\documentclass{nature}
\usepackage{amsmath}
\usepackage{amssymb}
\usepackage{lineno}
\usepackage{caption}
\usepackage{multibib}
\newcites{main,methods}{References,References}

\usepackage{graphicx}
\makeatletter
\let\saved@includegraphics\includegraphics
\AtBeginDocument{\let\includegraphics\saved@includegraphics}
\renewenvironment*{figure}{\@float{figure}}{\end@float}
\makeatother

\title{The Formation of Jupiter's Diluted Core by a Giant Impact}

\author{Shang-Fei Liu$^{1,2\star}$, Yasunori Hori$^{3,4}$, Simon M{\"u}ller$^{5}$, Xiaochen Zheng$^{6}$, \\
Ravit Helled$^{5}$, Doug Lin$^{7,8}$ \& Andrea Isella$^2$}

\begin{document}

\maketitle

\begin{affiliations}
 \item School of Physics and Astronomy, Sun Yat-sen University Zhuhai Campus, 2 Daxue Road, Tangjia, Zhuhai 519082, 
Guangdong Province, P.R. China;
 \item Department of Physics and Astronomy, Rice University, 6100 Main St., MS-108, Houston, TX 77005, USA; 
 \item Astrobiology Center, 2-21-1 Osawa, Mitaka, Tokyo 181-8588, Japan;
 \item National Astronomical Observatory of Japan, 2-21-1 Osawa, Mitaka, Tokyo 181-8588, Japan;
 \item Institute for Computational Science, Center for Theoretical Astrophysics and Cosmology, University of Zurich, Winterthurerstrasse 190, CH-8057 Zurich, Switzerland;
 \item Department of Physics and Center for Astrophysics, Tsinghua University, Beijing 10084, P.R. China;
 \item Department of Astronomy and Astrophysics, University of California, Santa Cruz, CA 95060, USA;
 \item Institute for Advanced Study, Tsinghua University, Beijing 100084, P.R. China.\\
 $^\star$To whom correspondence should be addressed; Email: liushangfei@mail.sysu.edu.cn
\end{affiliations}

\begin{abstract}
The \textit{Juno} mission\cite{2017SSRv..213....5B} is  designed to measure Jupiter's gravitational field 
with an extraordinary precision\cite{2017GeoRL..44.4694F}. Structure models of Jupiter that fit {\it Juno} 
gravity data suggest that Jupiter could have a diluted core and a total heavy-element mass $M_{\rm Z}$ ranging 
from ten to two dozens of Earth masses ($\sim 10-24 M_\oplus$). In that case the heavy elements are distributed 
within an extended region with a size of nearly half of Jupiter's radius $R_\textrm{J}$\cite{2017GeoRL..44.4649W,
2019ApJ...872..100D}. Planet formation models indicate that most of the heavy elements are accreted onto a compact 
core\cite{1996Icar..124...62P,2000ApJ...537.1013I,2014prpl.conf..643H}, and that almost no solids are accreted 
during runaway gas accretion (mainly hydrogen and helium, hereafter H-He), regardless to whether the accreted solids 
are planetesimals or pebbles\cite{2004A&A...425L...9P, 2010AJ....139.1297L, 2018A&A...612A..30B}. Therefore, 
the inferred heavy-element mass in the planet cannot significantly exceeds the core's mass. The fact that 
Jupiter's core could be diluted, and yet, the estimated total heavy-element mass in the planet is relatively large 
challenges planet formation theory. A possible explanation is erosion of the compact heavy-element core. 
Its efficiency, however, is uncertain and depends on both the immiscibility of heavy materials in metallic hydrogen 
and the efficiency of convective mixing as the planet evolves\cite{2004jpsm.book...35G, 2012ApJ...745...54W}. 
Neither can planetesimal enrichment and vaporization\cite{1982LPI....13..770S,2011MNRAS.416.1419H,2017ApJ...836..227L} 
produce such a large diluted core. Here we show that sufficiently energetic head-on collisions between 
additional planetary embryos and the newly emerged Jupiter can shatter its primordial compact core and mix 
the heavy elements with the outer envelope. This leads to an internal structure consistent with the diluted core scenario 
which is also found to persist over billions of years. A similar event may have also occurred for Saturn. 
We suggest that different mass, speed and impact angle of the intruding embryo may have contributed to 
the structural dichotomy between Jupiter and Saturn\cite{2005AREPS..33..493G,2012ApJ...750...52N,2013ApJ...767..113H}.
\end{abstract}

Giant impacts\cite{2010ApJ...720.1161L, 2015MNRAS.446.1685L} are likely to occur shortly after 
runaway gas accretion when a gas giant planet's gravitational perturbation significantly intensifies 
(about thirty-time increases in a fraction of a million years) and therefore destabilizes the orbits 
of nearby protoplanetary embryos. This transition follows oligarchic growth\cite{1998Icar..131..171K} 
and the emergence of multiple embryos with isolation mass in excess of a few 
$M_\oplus$\cite{2004ApJ...604..388I}. Some of these massive embryos may collide with the gas giant
during their orbit crossing\cite{2007ApJ...666..447Z,2013ApJ...775...42I}. Through tens of thousands 
of gravitational \textit{N}-body simulations with different initial conditions such as Jupiter's 
growth model, orbital configuration, etc.~(see Methods), we find that an emerging Jupiter has a strong influence on 
nearby planetary embryos. As a result, a significant fraction of these embryos could collide with Jupiter within a few 
million years, i.e., within the Solar nebula lifetime. Among those catastrophic events, head-on collisions are more 
common than grazing ones due to Jupiter's gravitational focusing effects. 

In order to investigate the influence of such impacts on the  internal structure of the young Jupiter we use the 
hydrodynamics code FLASH\cite{2000ApJS..131..273F} with the relevant equation of state (EOS). Details of the 
computational setup and the simulations are presented in the Methods section. Generally, the disintegration 
of the intruding embryo leads to the disruption of the planet's original core. However, to establish a large 
diluted-core structure as inferred from recent Jupiter structure models based on \textit{Juno}'s measurements, 
the core and embryos' fragments need to efficiently mix with the surrounding convective envelope, which requires 
a large embryo to strike the young Jupiter almost head-on. Massive embryos are available at the advanced stage 
of Jupiter's formation and our \textit{N}-body simulations also suggest that head-on collisions are common (see Methods).

In Figure 1 we show the consequence of a head-on collision between an embryo and Jupiter with 
an initial $M_{\rm core} = 10 M_\oplus$ silicate/ice core, a H-He envelope, approximately present-day total mass 
and radius (The young Jupiter may have a size up to twice of its current-day value, however, to avoid introducing 
additional free parameters, we consider models more similar to current Jupiter). In fact, the post-impact 
core-envelope structure mainly depends on the mass of the initial core and envelope as well as the impactor's mass 
and impact velocity $V_{\rm imp}$. We adopt an impact speed $V_{\rm imp} \sim 46$km s$^{-1}$ which is close to 
the free-fall speed onto Jupiter's surface (see Methods) and assume the impactor is comprised of an 8$\,M_\oplus$ 
silicate-ice core and a 2\,$M_\oplus$ H-He envelope. The total mass of proto-Jupiter's and embryo's core 
$M_\textrm{Z,total}$ is chosen to be compatible with the derived mass of heavy elements in Jupiter's diluted core 
models \cite{2017GeoRL..44.4649W}. Note that at Jupiter's distance of 5.2 AU from the sun, the impactor's speed 
relative to the gas giants is limited by the planets' surface escape speed. The acquisition of protoplanetary embryos 
would not lead to any major changes in the spin angular momentum and orientation of the targeted planet. The total 
energy injected into the young Jupiter by the intruding embryo is only a few percent of its original values so that 
there is little change in Jupiter's mean density and mass. 

The impact results in little mass loss (see Table 1), while Jupiter's initial core is completely disrupted. 
During the impactor's plunge and collision with the primordial core, a large amount of kinetic energy is dissipated. 
The heat release near the center increases the local temperature $T$, offsets the pressure $P$ balance, and induces 
oscillations (see the full video in Supplementary Information). The steep negative entropy gradient near the core 
overturns the local negative molecular weight gradient $\mu$ and leads to convection in the inner part of the envelope. 
Vigorous turbulence stirs up efficient mixing between the heavy elements and H-He envelope. After a few dynamical 
time-scales (a characteristic time scale to measure expansion or contraction of a planet; Jupiter's dynamical time scale 
is roughly a third of an hour), the initial silicate/ice core is thoroughly homogenized with the surrounding H-He and 
their mass fraction $Z \leq 0.5$ interior to $\sim 0.2$\,$R_\textrm{J}$. Within $\sim 30$ dynamical time-scales, 
Jupiter's interior settles into a quasi-hydrodynamic equilibrium with a diluted core extending to 
$R \sim 0.4-0.5 R_\textrm{J}$ (see Table 1 and panel a of Figure 2). In the 
outer half of the envelope, the gas density is slightly elevated and a small trace of the dredge-up heavy elements ($Z$) 
leads to the formation of a composition gradient.

The post-impact heavy-element distribution leads to a composition gradient that could evolve and become similar to 
an internal structure with a diluted-core. However, the hydrodynamic simulation is terminated ten hours after the impact. 
In order to explore under what conditions a diluted-core-like structure persists after the 4.56 Gyrs of Jupiter's 
evolution, we compute the thermal-evolution shortly after the impact until today. The hydro-simulation sets the initial 
heavy-element gradient as shown in panel a of Figure 2. Since the post-impact temperature profile 
is unknown\footnote{the exact temperature profile depends on the formation process\cite{Berardo2017,Cumming2018}, the 
energetics of the impact, etc.}, we consider various temperature profiles with different central temperatures. Furthermore, 
we consider an initial thermal structure that accounts for the accretion shock during runaway gas accretion as suggested 
by a recent Jupiter formation model\cite{Cumming2018} (see Methods for details). We find that for the head-on 
collision, a post-impact central temperature of $\sim$30,000 K leads to a current-state Jupiter with a diluted core. 
Another pathway to Jupiter's diluted core is if the initial temperature profile is shaped by the accretion shock. 
In panel b of Figure 2 we show the density profiles of the best-fitting models after the 4.56 Gyrs 
of evolution. If the central temperatures are higher (e.g., 50,000K), the interior is hot enough to "delete" the 
heavy-element gradient leading to a fully mixed planet. On the other hand, for low central temperatures ($\sim$ 20,000 K), 
convective mixing is less significant and the inferred density profile is less consistent with a diluted-core structure. 
Therefore, we conclude that Jupiter's diluted-core structure could be explained by a giant impact event, but under some 
specific conditions which include a head-on collision with a massive planetary embryo, a post-impact central temperature 
of $\sim$30,000 K, or an initial thermal structure created by the accretion shock during the runaway phase. Indeed, the 
hydrodynamic simulation implies that most of the impact energy is not deposited in the deep interior which results in 
lower central temperatures and to a diluted-core solution (see Methods).  

In contrast, if the same embryo collides with Jupiter at a grazing angle, it would be tidally disrupted gradually while 
sinking towards the center (see Figure 3). In Methods, we further show that impactors with Earth or 
sub-Earth mass disintegrate in gas giants' envelope before reaching their centers. Without smashing into the core directly, 
the shock wave induced by the impactor alone is insufficient to homogenize Jupiter's interior. These impacts generally 
lead to core growth rather than core destruction. Since impacts of planetary embryos are expected to be common after 
a gas giant's runaway gas accretion phase, a similar event with different impact conditions may have also happened to 
Saturn, and could in principle explain the dichotomy between the internal structures of Jupiter and 
Saturn\cite{2005AREPS..33..493G,2010ApJ...720.1161L,2012ApJ...750...52N,2013ApJ...767..113H}. 
A gradual accretion of planetesimals along with the runaway gas accretion may also 
produce a diluted core\cite{2017ApJ...836..227L,2017ApJ...840L...4H}.  A relevant issue to be investigate elsewhere 
is whether the steep compositional gradient needed to preserve the diluted core can also be established after a series 
of planetesimal-accretion events rather than a single embryo's giant impact. Finally, extra-solar gas giant planets could 
also suffer giant impacts which could explain some of the giant exoplanets with extremely large bulk 
metallicities\cite{2018AJ....155..214T}.

\textbf{Acknowledgements}  We thank S.M. Wahl and Y. Miguel for sharing their results with us. 
We thank J.J. Fortney, P. Garaud and H. Rein for helpful conversations. 
S.-F.L. thanks the support and hospitality from Aspen Center for Physics during 
the early stage of this work. 
D.L. thanks Institute for Advanced Study, Princeton, Institute of Astronomy and 
DAMTP Cambridge University for support and hospitality when this work was being 
completed. R.H. acknowledges support from SNSF grant 200021\_169054. 
A.I. acknowledges support from the National Aeronautics and Space Administration under
award No. 80NSSC18K0828 and from the National Science Foundation under grant No. AST-1715719.

 \textbf{Author Contributions} D.L. had the impact idea. S.-F.L. and A.I. examined its feasibility. 
 S.-F.L. coordinated this study. S.-F.L. and Y.H. designed and analysed the hydrodynamic simulations. 
 X.Z. and S.-F.L. performed and analysed the \textit{N}-body simulations. S.M. and R.H. designed 
 the long-term thermal evolution study. All authors contributed to discussions, 
 as well as editing and revising the manuscript. 
 
 \textbf{Author Information} The authors declare that they have no competing financial interests.
Correspondence and requests for materials should be addressed to S.-F.L. (email: liushangfei@mail.sysu.edu.cn).

\newpage

\begin{figure}
 \begin{center}
  \includegraphics[width=0.9\linewidth, clip=true]{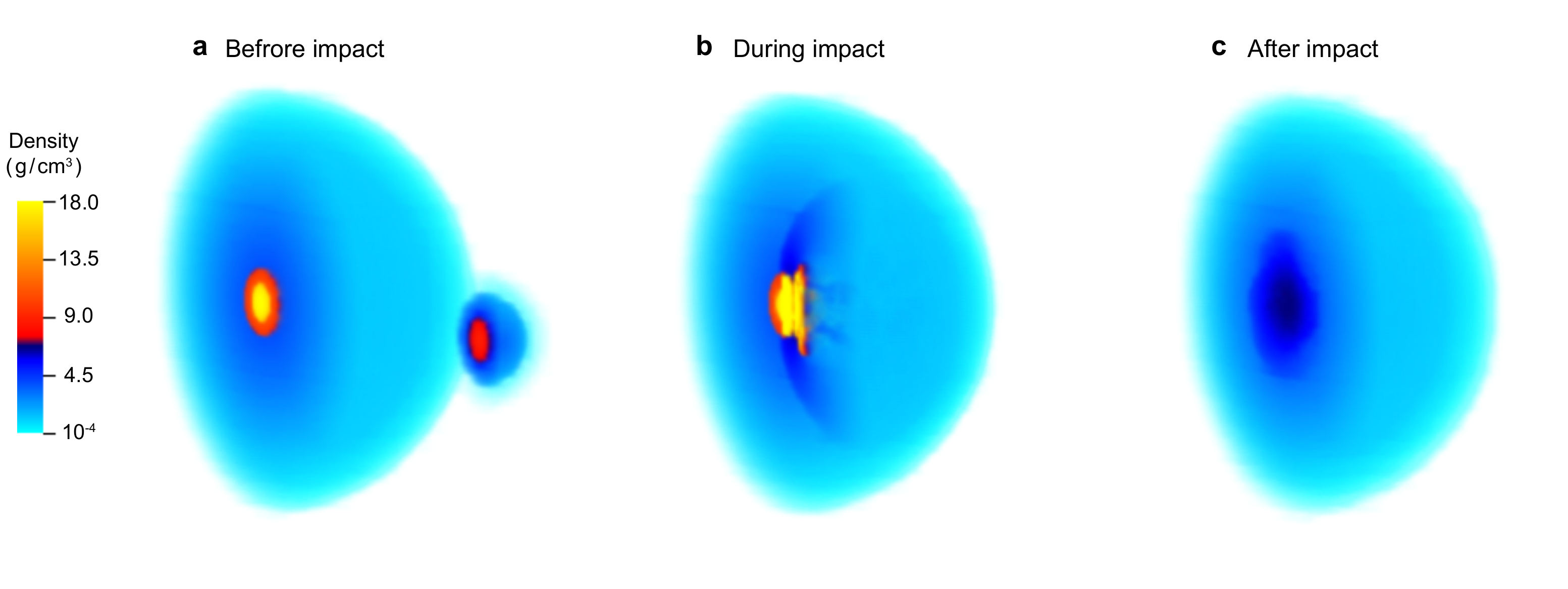}
  \caption{
   \textbf{3D cutaway snapshots of density distributions during a merger event between a proto-Jupiter
   with a 10\,$M_\oplus$ rock/ice core and a 10\,$M_\oplus$ impactor. 
   a,} just before the contact. \textbf{b,} the moment of core-impactor contact. 
  \textbf{c,} 10 hours after the merger. Due to impact-induced turbulent mixing, 
  density of Jupiter's core decreases by a factor of three after the merger, resulting in an extended diluted core. 
  A 2D presentation of density slices of the same event is shown in Extended Data Figure 3.
  }
 \end{center}
\end{figure}

\newpage
\begin{figure}
  \begin{center}
   \includegraphics[scale=0.65]{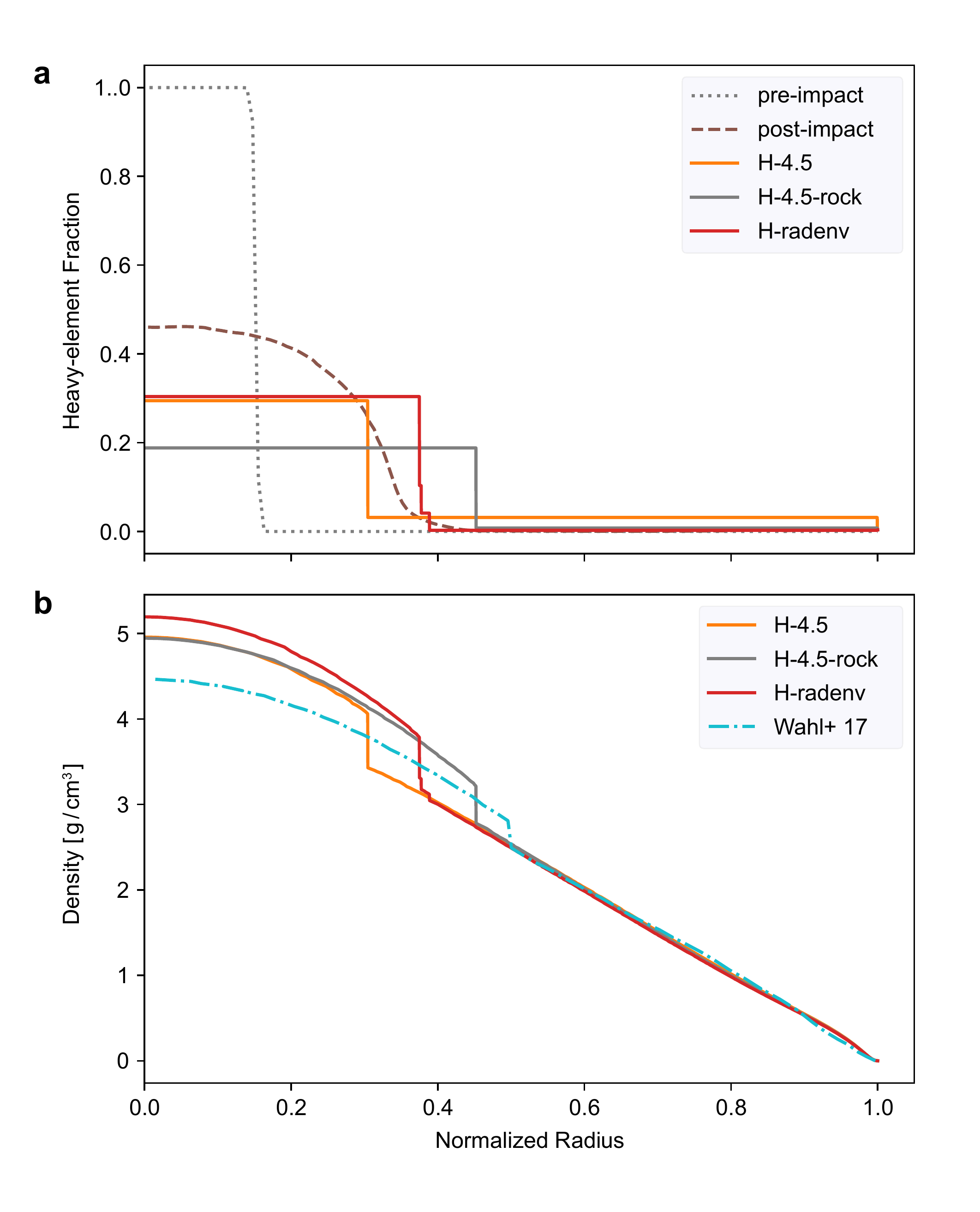}
   \caption{
  \textbf{Post-impact thermal evolution models. a,} Heavy-element 
  distribution vs. normalized radius before (dotted) and after (dashed) the giant impact. 
  The solid lines show the composition after 4.56 Gyrs of evolution for the three best-fit models 
  that result in a diluted core, see the Methods section and Table 3 for more details. 
  \textbf{b,} Density vs. normalized radius after 4.56 Gyrs of evolution (solid) and from the diluted-core 
  interior structure model of Wahl et al. 2017\cite{2017GeoRL..44.4649W}(dash-dotted). 
   }
  \end{center}
\end{figure}

\newpage
\begin{table*}
  \begin{center}
   \caption{\textbf{Initial conditions and final outcomes of the head-on giant impact simulation.}
  $M_\textrm{T}$ and $M_\textrm{I}$ are the total mass of the proto-Jupiter and the impactor, respectively. 
  $M_\textrm{core}$ is the mass of heavy elements in the proto-Jupiter's core. $M_{Z,\textrm{I}}$ and 
  $M_{Z,\textrm{total}}$ are the total mass of heavy elements contained in the impactor and the system, respectively.
  After the merger, values of total mass of Jupiter $M_\textrm{T}$ and heavy elements $M_{Z,\textrm{total}}$ are 
  measured within 1 and 2\,$R_\textrm{J}$, respectively. Those values reveal that the majority of Jupiter's mass 
  still resides in its original size, albeit a hot extended low-density envelope mostly made of H-He forms after 
  the merger (see also Extended Data Figure 3). The size of a diluted core was defined as the central 
  region enclosed by a sphere with $Z > 0.014$. The last three rows list values for the best-fit evolution models 
  to the interior structure model of Jupiter with a diluted core\cite{2017GeoRL..44.4649W}.
  $R_\textrm{core}/R_\textrm{J}$ is the radius of the proto-Jupiter's core scaled to the Jupiter's current radius.
  All mass quantities are in unit of $M_\oplus$.
  \label{tab:sims}}
  \begin{tabular}{c|cccccc} \hline
  & $M_\textrm{T}$ & $M_\textrm{core}$ & $M_\textrm{I}$ & $M_{Z,\textrm{I}}$ & $M_{Z,\textrm{total}}$ & $R_\textrm{core}/R_\textrm{J}$ \\ \hline \hline
  Before merger & 306.714 & 9.962 & 9.967 & 7.975 & 17.937 & 0.15 \\ 
  $\sim$10 hrs after merger & 304.946 / 313.360 & 17.693  & -- & -- & 17.901 / 17.925 & 0.423 \\  
  H-4.5: After 4.56 Gyrs & 313.36 & 10.61 & -- & -- & 17.925 & 0.30 \\
  H-radenv: After 4.56 Gyrs & 313.36 & 17.24 & -- & -- & 17.925 & 0.39 \\
  H-4.5-rock: After 4.56 Gyrs & 313.36 & 15.92 & -- & -- & 17.925 & 0.45 \\ \hline
  \end{tabular}   
  \end{center}
\end{table*}

\newpage
\begin{figure}
  \begin{center}
   \includegraphics[width=0.7\linewidth,clip=true]{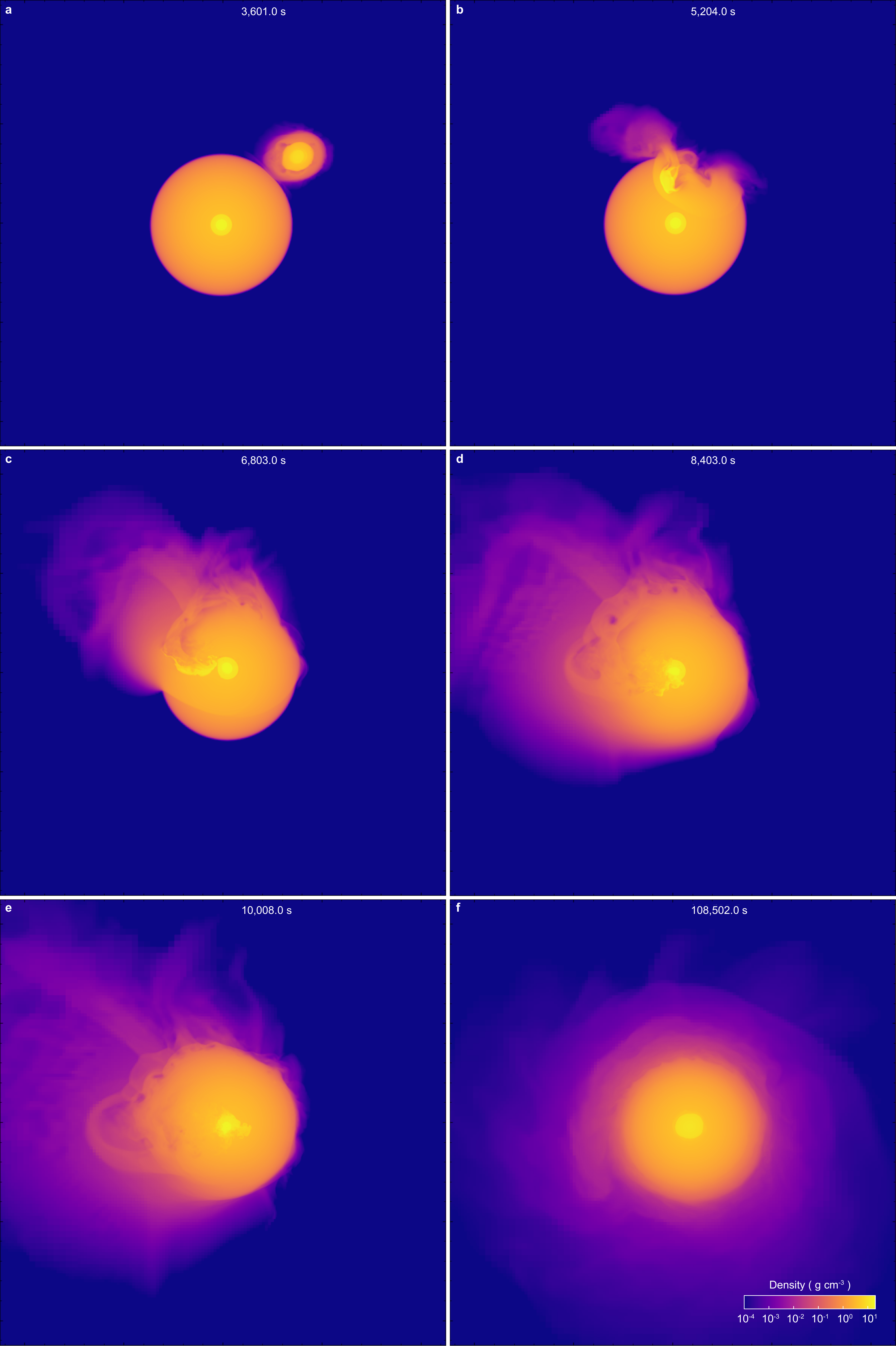}
   \caption{
\textbf{2D snapshots of an off-center collision between the proto-Jupiter with 
  a 10\,$M_\oplus$ solid core and a 10\,$M_\oplus$ impactor. a,}  Density contours in the 
  orbital plane before the impact; \textbf{b-e,} the impactor being disrupted and accreted; 
  \textbf{e,} at $\sim$30 hours after the impact. See Methods for detailed discussion.
   }
  \end{center}
\end{figure}
\clearpage

\section{Methods}

\subsection{A statistical \textit{N}-body study of embryo collisions:}
We investigate the statistics of collisions between an emerging Jupiter and planetary embryos with the open-source 
\textit{N}-body code REBOUND\cite{2012A&A...537A.128R} version 3.6.2. To simulate the evolution of a planetary 
system we choose the built-in hybrid HERMES integrator\footnote{In recent updates of REBOUND, the HERMES integrator 
has been replaced by the MERCURIUS integrator, which offers a similar capability in a single scheme.}, which uses 
the WHFast integrator\cite{2015MNRAS.452..376R} for the long-term dynamics and switches to the IAS15 
integrator\cite{2015MNRAS.446.1424R} when close encounters (such as scattering and collisions) happen. 

Our \textit{N}-body simulations start from a coplanar configuration in which five ten-Earth-mass planetary 
embryos ($M_\textrm{p} = 10 M_\oplus$) orbit the Sun ($M_* = 1 M_\odot \sim 3.3\times 10^5 M_\oplus$) on circular 
prograde orbits. The embryo at 5.2 astronomical units (a.u.) from the Sun grows into a Jupiter-mass planet 
at the end of the simulation. Initially, two embryos are placed interior to Jupiter's orbit and the other 
two embryos are placed exterior to Jupiter's orbit. The orbital separation between two adjacent embryos 
$i$ and $i+1$ is determined by a dimensionless number  
\begin{equation}
  k = \frac{2a_i}{r_\textrm{H}} \left(\frac{a_{i+1} - a_{i}}{a_{i+1}+a_{i}}\right)\;,
\end{equation}
where $a_i$ and $a_{i+1}$ are the semi-major axes of each embryo, and 
$r_\textrm{H} = a_i \left(\frac{M_\textrm{p}}{3M_*}\right)^{1/3}$ is the Hill radius of embryo $i$. It is 
convenient to express Equation 1 in terms of $q=a_{i+1}/a_i$, the ratio of semi-major axis between embryos 
$i$ and $i+1$ 
\begin{equation}
  k = 2 \left(\frac{3}{\mu}\right)^{1/3} \left(\frac{q - 1}{q + 1}\right)\;,
\end{equation}
where $\mu=M_\textrm{p}/M_* \simeq 3\times 10^{-5}$ is the mass ratio between the embryo and the Sun. A larger 
$k$ will give rise to a wider separation, i.e. a more dynamically stable configuration. Extended Data Table 1 
summarizes the locations of all embryos for a given parameter $k$ in our \textit{N}-body simulation suite. 
In addition, we also consider a configuration, in which all four embryos are beyond Jupiter's orbit. 

At the onset of the simulation, the runaway gas accretion of Jupiter's core starts. The mass accretion rate 
is an exponential decay function characterized by an exponential time parameter $t_\textrm{grow}$ ranging from 
0.1 Myr to 0.5 Myr in this study. At a given time $t$, the mass of an emerging Jupiter is determined by
\begin{equation}
  M(t) = M_\textrm{J} - (M_\textrm{J}-M_\textrm{p}) \; e^{-t/t_\textrm{grow}},
\end{equation}
where $M_\textrm{J} = 317.8 \,M_\oplus $ is one Jovian mass. In this model, Jupiter quickly acquires more 
than 90\% of its mass within 3 $t_\textrm{grow}$ and steadily gains another a few percent of its mass until 
$t = 10\;t_\textrm{grow}$. For simplicity, we assume that all other four embryos do not grow during the whole time, 
since a typical hydrostatic growth stage of an embryo before it entering the runaway gas accretion phase
is around a few Myr and the embryo mass barely increases. 

Size is another crucial factor as a larger cross-section can boost the probability of collisions. 
We adopt the mean density of the Earth for embryos, so their sizes $R_\textrm{p} \simeq 2.15R_\oplus$, where 
$R_\oplus$ is Earth's mean radius. For the emerging Jupiter, its mean density could be as low as half of its 
current-day value. We use the parameter $f$ to describe the degree of inflation. 

Thus, we design a simple classification for our \textit{N}-body simulation suite with three free parameters 
$k$, $t_\textrm{grow}$ and $f$. For each combination set of ($k$, $t_\textrm{grow}$, $f$), we run thousands 
of simulations with other orbital parameters (e.g., true anomaly, argument of periapsis) randomly chosen 
between 0 and $2\pi$. 

At the end of an \textit{N}-body simulation ($t=10\;t_\textrm{grow}$), a planetary embryo may remain bound to 
the Sun with considerable changes in its orbit, or coalesce with Jupiter and other embryos or escape from the 
system after a close encounter. The statistics of final outcomes of four planetary embryos under the influence 
of an emerging Jupiter is shown in Extended Data Figure 1. The results are grouped by different parameters to 
compare their impacts. In all subsets of our \textit{N}-body simulations, we observe an efficient pathway to 
deliver planetary embryos to collide with an emerging Jupiter.

Because embryos are equally distributed on both sides of Jupiter's orbit (except for the last group with 
all embryos in the "Outward" state to begin with), the results suggest that embryos both interior or exterior to 
Jupiter could collide with Jupiter within the simulation time. While embryos beyond Jupiter may have a 
slightly larger chance to strike Jupiter as there are less embryos asymptotically result in an "Outward" destiny. 
Among the three key parameters, orbital tightness characterized by $k$ plays the most substantial role 
in affecting the collision probability. For the same orbital configuration, Jupiter inflation factor $f$ 
can slightly change the collision rate. However, Jupiter's accretion history determined by $t_\textrm{grow}$ 
has the least influence on the results.  

We analyze the distribution of collision angle using our \textit{N}-body simulation suite. And the histograms 
of collision angles are plotted in Extended Data Figure 2. Each histogram represents a detailed breakdown 
of "Merger" events of a simulation set presented in Extended Data Figure 1. Unlike collisions between 
similar-sized planetary bodies, in which 45$^\circ$ collisions are common\cite{2006Natur.439..155A}, 
the statistical results suggest that half of the merger events have collision angles less than $\sim $ 30$^\circ$ 
in all cases we investigated. We suggest that low-angle impacts are very common because of Jupiter's strong 
gravitational focusing effect.

It is often useful to define a two-body escape velocity as
\begin{equation}
    V_\textrm{esc}= \left(\frac{2 \textrm{G}(M_\textrm{J}+M_\textrm{p})}{R_\textrm{R}+R_\textrm{p}}\right)^{1/2},
\end{equation}
which is around 51 km s$^{-1}$ for the proto-Jupiter and the 10 $M_\oplus$ impactor studied in the hydrodynamic 
simulation. In general, an embryo's impact velocity $V_\textrm{imp}$ is related to $V_\textrm{esc}$ as well as 
the local Keplerian velocity $V_\textrm{kep}$. Gravitational perturbation during close encounters can produce 
an impact velocity with a magnitude up to the escape velocity\cite{1997ApJ...477..781L}. On the other hand, 
the Keplerian orbital velocity gives rise to the random velocity dispersion during impacts. At Jupiter's current 
location, $V_\textrm{kep} \sim$ 13 km s$^{-1}$ is much smaller than $V_\textrm{esc}$, so the impact velocity 
$V_\textrm{imp}$ is approximately at the escape velocity $V_\textrm{esc}$. Indeed, we find the impact velocity 
is quantitatively similar to $V_\textrm{esc}$ rather than $V_\textrm{kep}$, although $V_\textrm{imp}$ is always 
slightly smaller than $V_\textrm{esc}$ in the \textit{N}-body simulation suite, because initial separations between 
Jupiter and embryos are finite (a two-body system has a negative gravitational potential energy). 

This simple statistical model can be improved in the future to compare with other formation models of the outer 
Solar system. For example, because Jupiter's inward migration is much slower than those planetary embryos, the 
presence of Jupiter in the Solar nebula acts like a barrier for inward migrating planetary embryos formed exterior to 
Jupiter\cite{2015ApJ...800L..22I}. Consequently, collisions among those planetary embryos may become frequent 
and some of those events may eventually form Uranus and Neptune\cite{2015A&A...582A..99I}.

\subsection{Hydrodynamic simulations:}
Our 3D hydrodynamic simulation of giant impacts between a proto-Jupiter and a protoplanetary embryo 
is based on the framework of the Eulerian FLASH code\cite{2000ApJS..131..273F} which utilizes the 
adaptive-mesh refinement. The setup of giant impact simulations has been laid out in our previous 
study\cite{2015ApJ...812..164L}. Here, we briefly describe the model of the planetary interior. 
The primordial Jupiter is modeled with a three-layer structure: a silicate core, an icy mantle, and a 
H-He envelope. We calculate two thermodynamic (density and internal energy) properties of silicate and 
ice material and their velocities with the governing continuity, momentum, and energy equation. For 
computational efficiency, these quantities are converted into pressure and temperature with the Tillotson 
EOS\cite{Melosh:1989uq}.  The mass fraction between ice to silicate is assumed to be 2.7 according 
to that of protosun (2--3). In addition, the H-He EOS is modeled with an $n = 1$, $\gamma = 2$ polytropic 
relation, where $n$ and $\gamma$ are the polytropic and adiabatic indexes. Although this idealized 
treatment ignores effects such as the H-He phase transition and separation, it reasonably matches the 
density profile of Jupiter's envelope calculated with {\it ab-initio}  EOS\cite{2013ApJ...762...37L}, 
and is good enough for dynamic processes that happen in a few hours (see detailed discussion below).

\textit{Collisions between a proto-Jupiter with a 10 $M_\oplus$ core and a 10 $M_\oplus$ embryo:} 
From \textit{N}-body simulations we learn that most collisions have collision angles less than 30 degrees, 
so we first study the head-on collision as one of the representative cases in the main text and the consequence 
is shown in Figure 1. Here we also plot its 2D counterpart in Extended Data Figure 3. The general 
behavior of head-on collisions has been studied extensively in previous 
works\cite{2015MNRAS.446.1685L}$^{,}$\cite{2015ApJ...812..164L}. 
To recapitulate, the solid material of the impactor can penetrate Jupiter's gaseous envelope and smash into 
its core as a whole. As a result, Jupiter's core gets completely destroyed after the impact. The release of 
a large amount energy inside the proto-Jupiter drives large scale turbulence and the primordial compact core 
is homogenized subsequently. We compare the enclosed internal energy of Jupiter as a function of radius before 
and after the impact. The results are shown in Extended Data Figure 4. Although Jupiter gains internal energy 
through the release of kinetic and gravitational energy of the impactor as well as impactor's own internal energy, 
the core region gets barely heated. In fact, there is even a little decrease of internal energy inside the core 
region right after the impact possibly due to mixing with H-He. The analysis suggests that the impactor dumps most 
of its energy outside the original core region. 

Our simplified EoS for H/He causes less efficient dissipation of the impactor’s kinetic energy within the H/He 
envelope. As a vigorous mixing between H/He and core material, however, is driven by a merger between the core 
of a photo-Jupiter and an impactor, we can expect formation of a dilute core to occur regardless of EoS models. 
In addition, a temperature profile inside a core is not strongly affected by the choice of a H/He EoS model 
because the impact causes only a small change in internal energy inside the core.

To illustrate the effects of off-center collisions, we run the same setup of simulation except that 
the collision angle is at 45 degrees. The consequence is shown in Figure 3. 
Because the initial impact velocity is at the escape velocity, the impactor misses Jupiter's core and 
overshoots until Jupiter's gravitational force pulls it back. During its course, the impactor gradually 
loses angular momentum and gets torn apart. The remnant is gently accreted by Jupiter's solid core later on. 
As a result, the impact has little influence on Jupiter's core-envelope structure.   

\textit{A head-on collision between a proto-Jupiter with a massive core and a small impactor:}
In addition, we perform a head-on collision between a proto-Jupiter with a massive primordial core of 
17\,$M_\oplus$ and a 1 $M_\oplus$ impactor, which is composed of pure silicate, at the same impact velocity. 
The total amount of heavy elements is the same as that in previous head-on and off-center models (hereafter, 
case-1 and case-2). Unlike case-1, the impactor disintegrates in the proto-Jupiter's envelope before making 
contact with the core. A strong shock wave induced by the entry of the impactor propagates throughout the 
entire planet and deforms the core (see panel c of Extended Data Figure 5). The heavy elements are well mixed 
with a small fraction of H-He (only $\sim$5wt\%) inside the proto-Jupiter's core after the impact because of 
a weak impact-induced oscillation and less efficient turbulent mixing. As a result, the central density of the 
core still decreases by a factor of two thirds. Although the core--envelope boundary slightly spreads out, a steep 
density gradient between the core and H-He envelope is preserved, leading to the retention of a compact, massive core.

To summarize, only in case-1 we observe a smooth transition between the core and H-He envelope after the 
impact, as the impactor is massive and hits Jupiter's core directly. However, in both case-2 and case-3, 
because the impactor is unable to collide with the core as an integrated body, the proto-Jupiter's core
becomes less chemically homogenized after it gets restored from deformation. Therefore, we conclude that 
neither a small impactor, nor an off-center collision is able to form a large diluted core, and proto-Jupiter 
with a primordial solid core should have experienced a catastrophic nearly head-on collision with a large 
embryo, if the present-day Jupiter has a massive, diluted core. A more comprehensive parameter study, including 
a range of impactor's mass and speed as well as off-center collisions, will be presented elsewhere.

\subsection{Post-impact thermal evolution:}
We simulate Jupiter's long-term evolution after the giant impact in order to identify the evolutionary paths 
that lead to a diluted core structure at present-day. The planetary evolution is modelled using the 1D stellar 
evolution code Modules for Experiments in Stellar Astrophysics (MESA), where the planet is assumed to be 
spherically symmetric and in hydrostatic equilibrium \cite{Paxton2011, Paxton2013, Paxton2015, Paxton2018}. 
The evolution is modeled with a modification to the equation of state\cite{Mueller2019}, where the H-He EOS 
is based on SCVH\cite{Saumon1995} with an extension to lower pressures and temperatures, and the heavy-element 
(H$_2$O/SiO$_2$) EOS is QEOS\cite{More1988, Vazan2013}. Conductive opacities are from Cassisi et al.
(2007)\cite{Cassisi2007}, and the molecular opacity is from Freedman et al. (2007)\cite{Freedman2007}.

The planetary evolution is governed by the energy transport in the interior, which can occur via radiation, 
conduction, or convection. We use the standard Ledoux criterion\cite{Ledoux1947} to determine whether 
a region with composition gradients is stable against convection, i.e., $\nabla_{T} < \nabla_{ad} + B$, where 
$\nabla_{T} = d \log T / d \log P$, with $\nabla_\textrm{ad}$ and $B$ being the adiabatic temperature and 
composition gradient, respectively. If the composition gradient is such that the mean molecular weight increases 
towards the planetary center, then $B > 0$ and the composition gradient could inhibit convection. For a homogeneous 
planet, $B = 0$ and the Ledoux criterion reduces to the Schwarzschild criterion $\nabla_{T} < \nabla_\textrm{ad}$. 
A region that is Ledoux stable but Schwarzschild unstable could develop semi-convection. In that case, 
double-diffusive processes can lead to additional mixing\cite{2011ApJ...731...66R}.

In the planet evolution code, convective mixing is treated via the mixing length theory (MLT), which provides 
a recipe to calculate  $\nabla_{T}$ and the diffusion coefficient, fully determining the convective flux. 
The MLT requires the knowledge of a mixing length $l_\textrm{m} = \alpha_\textrm{mlt} H_{P}$, where $H_{P}$ is 
the pressure scale-height and $\alpha_\textrm{mlt}$ is a dimensionless parameter. The expected value of 
$\alpha_\textrm{mlt}$ for planets is poorly constrained. Following previous work on Jupiter's evolution with 
convective mixing\cite{2018A&A...610L..14V} we use $\alpha_\textrm{mlt} = 0.1$ as our baseline. 
It is found that the mixing is relatively insensitive to the choice of the mixing length within about 
an order of magnitude. This is because its value  does not directly determine when mixing occurs, but 
the mixing efficiency. To investigate the sensitivity of the results on this parameter we also included 
a model with $\alpha_\textrm{mlt} = 10^{-3}$. While our conclusions on the diluted core are robust, a detailed 
and rigorous investigation on mixing in giant planets is clearly desirable, and will be presented in future 
work\cite{Mueller2019}.

The case of semi-convection is treated as a diffusive processes\cite{Langer1983} which requires the 
calculation of the temperature gradient and diffusion coefficient in the semi-convective region. 
The recipe includes a free parameter that can be interpreted as the layer-height of the double-diffusive 
region\cite{Wood2013,Radko2014}, which is unknown and could range over a few orders of magnitude. 
In the case where we include semi-convection, we set the value to $10^{-5}$ pressure scale heights, which 
is an intermediate value in the range given in the literature\cite{2012A&A...540A..20L}.

The hydro-simulation of the giant impact sets the post-impact composition profile to be used by the evolution 
model. The initial temperature profile is crucial for determining the energy transport for the subsequent evolution. 
Since proto-Jupiter's thermal state at the time of impact is unknown, we consider various initial temperature 
profiles and explore how the mixing is affected by this choice. Giant planet formation calculations estimate the 
central temperature of proto-Jupiter to be  $\sim 10^4$ K\cite{Cumming2018}. The exact temperature, however, 
is unknown and can change by a factor of a few. For determining the convective mixing efficiency such factors can 
lead to large differences in the long-term evolution and the final internal structure. Also, recent work has shown 
that accounting for the accretion shock during the runaway gas accretion phase can lead to a radiative envelope and 
a non-monotonic temperature profile in the deep interior\cite{Berardo2017,Cumming2018}. We include this 
possibility in one of our models (H-radenv). Our nominal models use $\alpha_\textrm{mlt} = 0.1$, no semi-convection 
with the heavy elements being represented by water. A summary of the model parameters is given in Extended Data Table 2. 

In Extended Data Figure 6 we present the starting models that are evolved to Jupiter's age. The solid and dashed lines 
correspond to the head-on and oblique (at an angle of 45 degrees) collisions, respectively. The temperatures are 
increasing towards the interior for all models except H-radenv, as explained above. Here, a temperature inversion occurs 
in the deep interior, corresponding to the location of the accretion shock during early runaway gas accretion. Note 
that in this model the location of the temperature-inversion occurs within the same region of the composition gradient, 
which supports the stability of the region against convection. While the exact location of the temperature jump is not 
well determined, it can be estimated due to the requirement of reaching the so called cross-over mass to enter the 
runaway phase\cite{2014prpl.conf..763B}. As the heavy-element fraction increases, the interior becomes hotter 
due to the change in opacity and the increase in density. If the collision is head-on, the composition gradient is 
shallower and extends farther into the envelope. 

Extended Data Figures 7 \& 8 show the density profiles after 4.56 Gyrs of evolution for the head-on and oblique 
collision, respectively. The crucial influence of the initial thermal profile on the mixing is clear: For the 
$\log T_\textrm{c} [\textrm{K}] = 4.7$ head-on collision case (H-4.7), the end-result is a fully homogeneous 
Jupiter without a core. For the oblique impact, even the very steep composition gradient, with the highest 
temperatures, is insufficient to inhibit substantial mixing of the deep interior. The intermediate temperature 
profiles lead to varying degrees of mixing. In general, the head-on collision results in an extended core that 
is highly enriched in H-He, while for the oblique impact the core is more compact and less diluted. Despite a 
substantial fraction of proto-Jupiter being very hot in the model H-radenv, there is not enough mixing to erase 
the composition gradient. In this case, the envelope is radiative at early times when mixing would be most 
efficient. If a lower mixing length is chosen (H-4.5-low$\alpha$), the composition gradient is less eroded and 
extends farther into the envelope. Because the energy transport is also affected by the chosen mixing length, 
Jupiter's interior is hotter and denser compared to H-4.5. 

In H-4.5-semiconv, we consider the same model as H-4.5 but allow semi-convective mixing. with a layer height 
of $10^{-5}$ pressure scale-heights. In this case, semi-convection is insufficient to overcome the stabilizing 
composition gradient. While some additional mixing occurs, particularly at early times, there are no semi-convective 
regions towards the end of the evolution. In other words, the final interior structure is such that the radiative 
regions are Schwarzschild and Ledoux stable. This demonstrates that also when semi-convection is included we infer 
a Jupiter with a diluted core.

In order to completely erase the composition gradient created by the giant impact the impact must be head-on with 
a very hot interior ($\sim$ 50,000 K) with the heavy elements represented by water (H-4.7). In all the other models 
we consider, the stabilizing effect of the post-impact heavy-element distribution is inhibiting the development 
of convective instabilities  resulting in an inhomogeneous Jupiter. Therefore, the typical outcome of the 
calculation is an interior structure that is not fully mixed and is characterized by several radiative-convective 
interfaces. Interestingly, the development of these interfaces seems to be a common occurrence when modelling 
Jupiter's evolution with composition gradients\cite{2018A&A...610L..14V,Mueller2019}. If the core is defined 
as the region that is heavy-element rich in comparison to the envelope, then most of our models imply that Jupiter 
has a diluted and extended core extending to $\sim 30\%-50\%$ of the planet's radius. All the oblique collisions 
lead to a relatively compact core since the initial composition gradient is very steep. 
 
Figure 2 shows the models that best match the diluted-core density profile from 
Wahl et al. (2017)\cite{2017GeoRL..44.4649W} (H-4.5-rock, H-4.5, H-radenv). We find that for the head-on 
collision, a post-impact central temperature of $\sim$ 30,000 K leads to a current-state Jupiter with a diluted 
core (H-4.5 and H-4.5-rock). If the heavy elements are represented by rock (SiO$_2$), the diluted core extends 
farther into the envelope and is thus more consistent with a Jupiter structure with a diluted core. Another pathway 
to the diluted core is when Jupiter's deep interior is radiative due to the accretion shock as predicted by recent 
giant planet formation models\cite{Cumming2018} (H-radenv). Videos that demonstrate the planetary evolution for three 
selected cases can be found in the Supplementary Information.

\textbf{Data availability.}
The datasets generated and analysed during the current study 
are available from the corresponding authors upon reasonable request.

\textbf{Code availability.}
The FLASH code is publicly available for download at http://flash.uchicago.edu/site/flashcode. The implementation of giant 
impact simulations in the framework of FLASH is available upon request. The REBOUND code is publicly available for download 
at https://github.com/hannorein/rebound. The MESA code is an open source stellar evolution code and is publicly available 
at http://mesa.sourceforge.net. The modified version of the MESA code is not yet ready for public release - it will be 
presented in future work\cite{Mueller2019}. Gnuplot, Jupyter Notebook, Mathematica, VisIt and yt python package 
have been used for data reduction and presentation in this study.

\clearpage

\begin{figure}[htbp]
  \begin{center}
    \includegraphics[width=0.75\linewidth,clip=true]{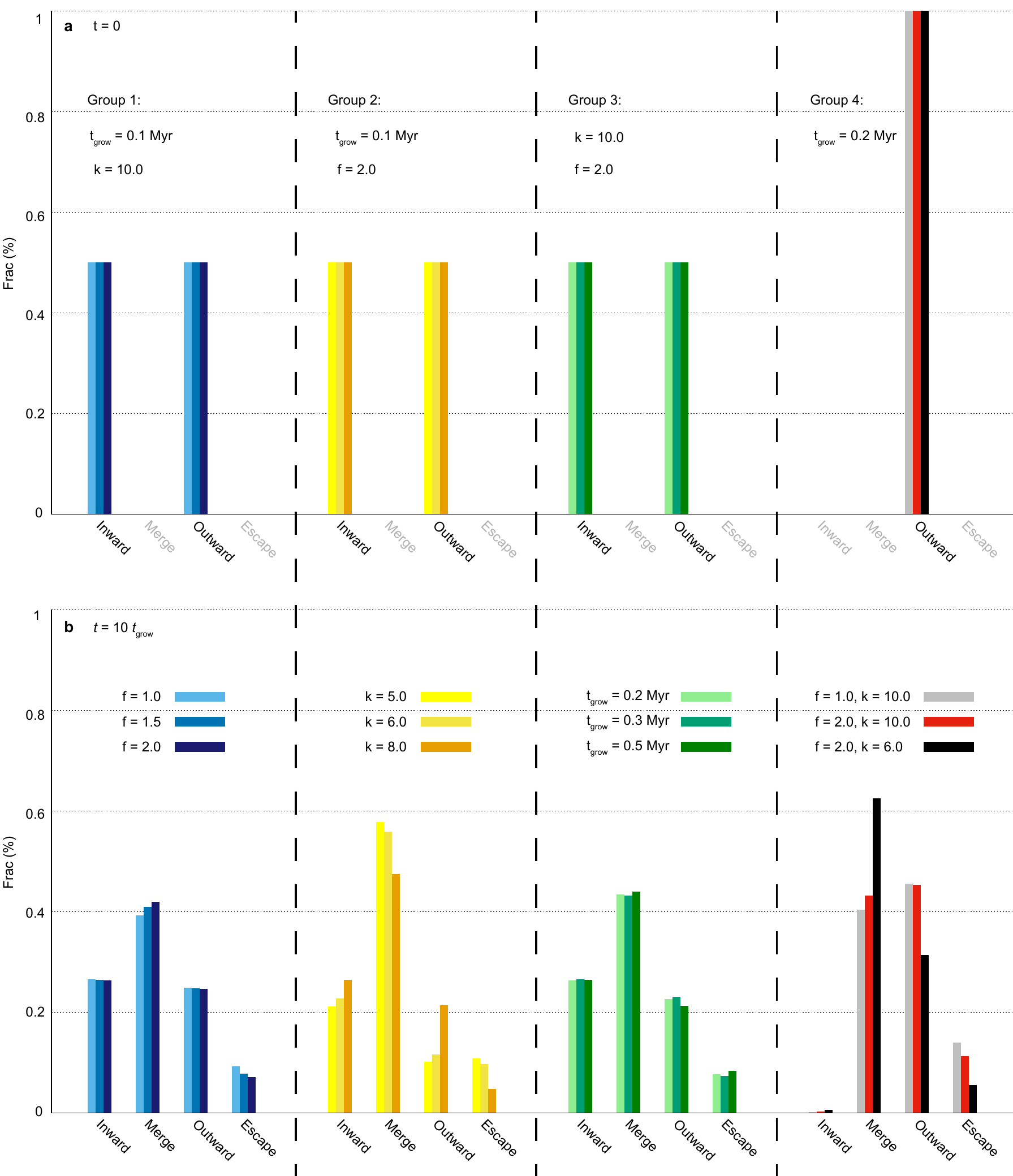}
    \caption{
\textbf{Extended Data Figure 1 | Statistics of outcomes of four planetary embryos under the influence of an emerging Jupiter. 
  a,} the initial configurations of four planetary embryos divided into four groups based on fixed parameters 
  shown under the group numbers. In group 1-3, half of the embryos are placed inside Jupiter's orbit (labeled as 
  "Inward"), the other half are outside Jupiter's orbit (labeled as "Outward"). In group 4, all embryos are 
  outside Jupiter's orbit. The exact location of every embryo is shown in Table 1 in the Supplementary Information. 
  \textbf{b,} the statistical outcomes of the dynamic evolution after 10 $t_\textrm{grow}$. Because Jupiter's 
  growth can substantially modify orbits of those embryos. Some embryos collided with Jupiter (labeled as "Merger"), 
  and some have been ejected from the Solar system (labeled as "Escape"). Other embryos are labeled either "Inward" 
  or "Outward" depending on their orbital locations inside or outside Jupiter's orbit. Colours indicate different 
  choices of the free parameter displayed in legend in each group.
    }
  \end{center}
\end{figure}

\begin{figure}[htbp]
  \begin{center}
    \includegraphics[width=0.85\linewidth,clip=true]{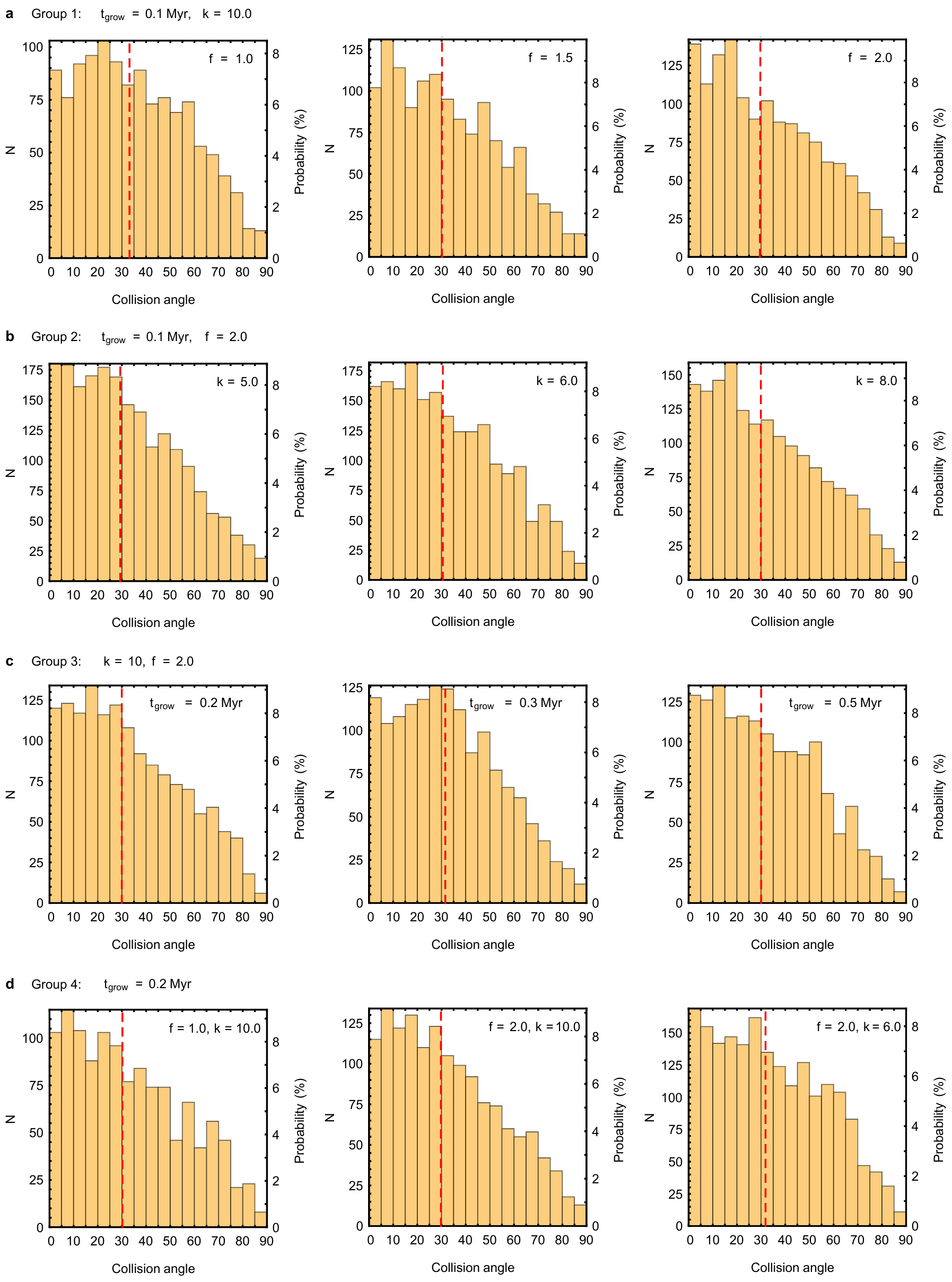}
    \caption{
\textbf{Extended Data Figure 2 | Histograms of collision angles of each data set presented in 
  Extended Data Figure 1. a,} group 1. \textbf{b,} group 2. \textbf{c,} group 3. \textbf{d,} 
  group 4. The bin size is 5$^\circ$, and there are 18 bins in each plot. 
  The red dashed line indicate the median value in each case. The results suggest head-on collisions are 
  more common than grazing ones.
    }
  \end{center}
\end{figure}
   
\begin{figure}[htbp]
  \begin{center}
    \includegraphics[width=\linewidth,clip=true]{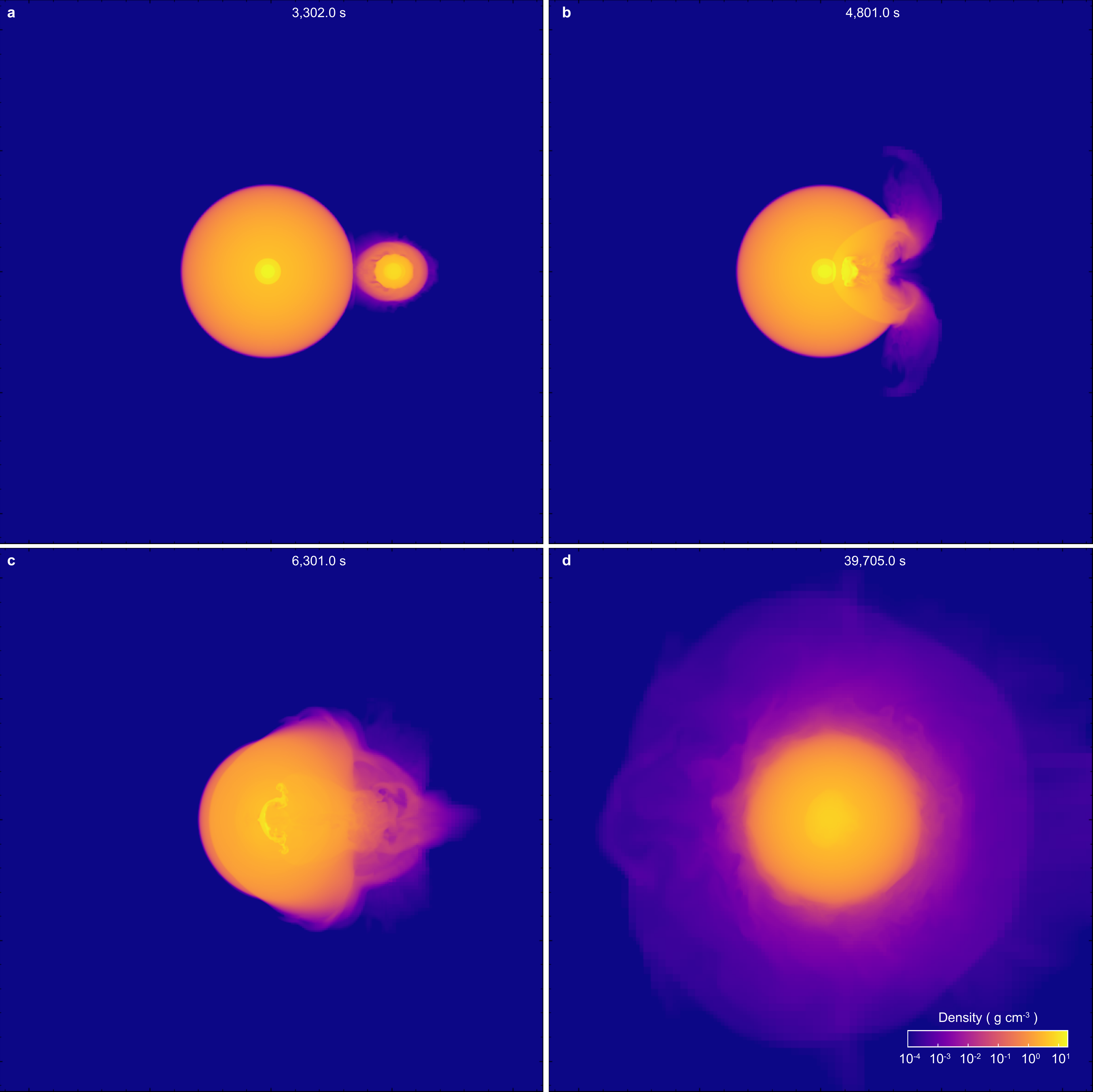}
    \caption{
\textbf{Extended Data Figure 3 | 2D snapshots of a merger between the proto-Jupiter with 
  a 10\,$M_\oplus$ solid core and a 10\,$M_\oplus$ impactor. a,}  
  Density contours in the orbital plane before the impact; \textbf{b,} before the impactor arriving at the core; 
  \textbf{c,} after the destruction of the core; \textbf{d,} at $\sim$10 hours after the impact.
    }
  \end{center}
\end{figure}   

\begin{figure}[htbp]
  \begin{center}
    \includegraphics[width=0.8\linewidth,clip=true]{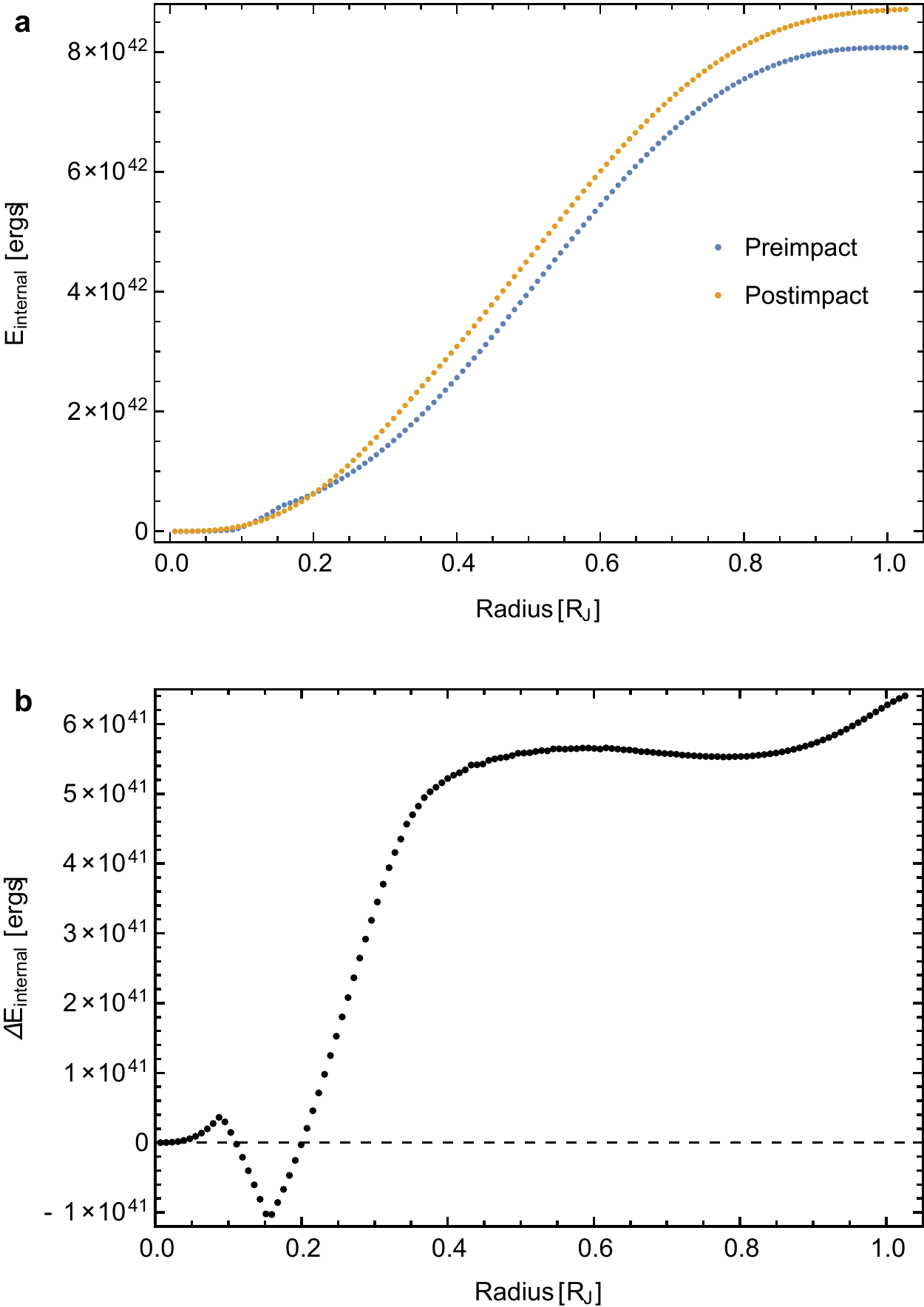}
    \caption{
\textbf{Extended Data Figure 4 | The change of internal energy caused by the merger in case-1. a,} 
  The enclosed internal energy of Jupiter before and after the impact as a function of 
  radius. \textbf{b,} The net change of enclosed internal energy of Jupiter as a function of radius.
    }
  \end{center}
\end{figure}   
   
\begin{figure}[htbp]
  \begin{center}
    \includegraphics[width=\linewidth,clip=true]{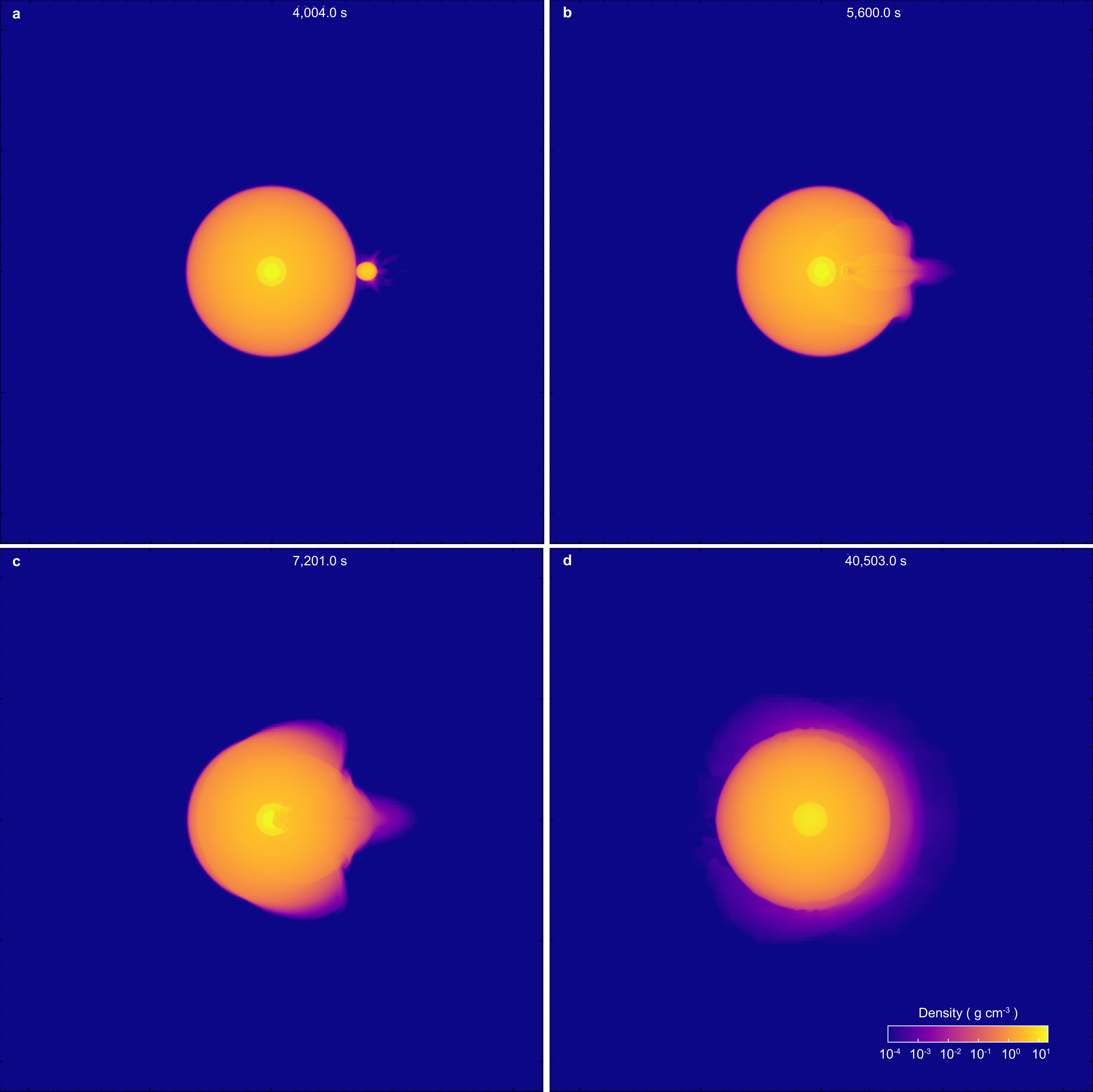}
    \caption{
\textbf{Extended Data Figure 5 | 2D snapshots of a merger between the primordial Jupiter with 
  a 17\,$M_\oplus$ core and a 1\,$M_\oplus$ impactor. a,} 
  Density contours in the orbital plane before the impact; \textbf{b,} before 
  the impactor arriving at the core; \textbf{c,} after the merger with the core; 
  \textbf{d,} at $\sim$10 hours after the impact.
    }
  \end{center}
\end{figure}   
   
\begin{figure}[htbp]
  \begin{center}
    \includegraphics[width=0.8\linewidth,clip=true]{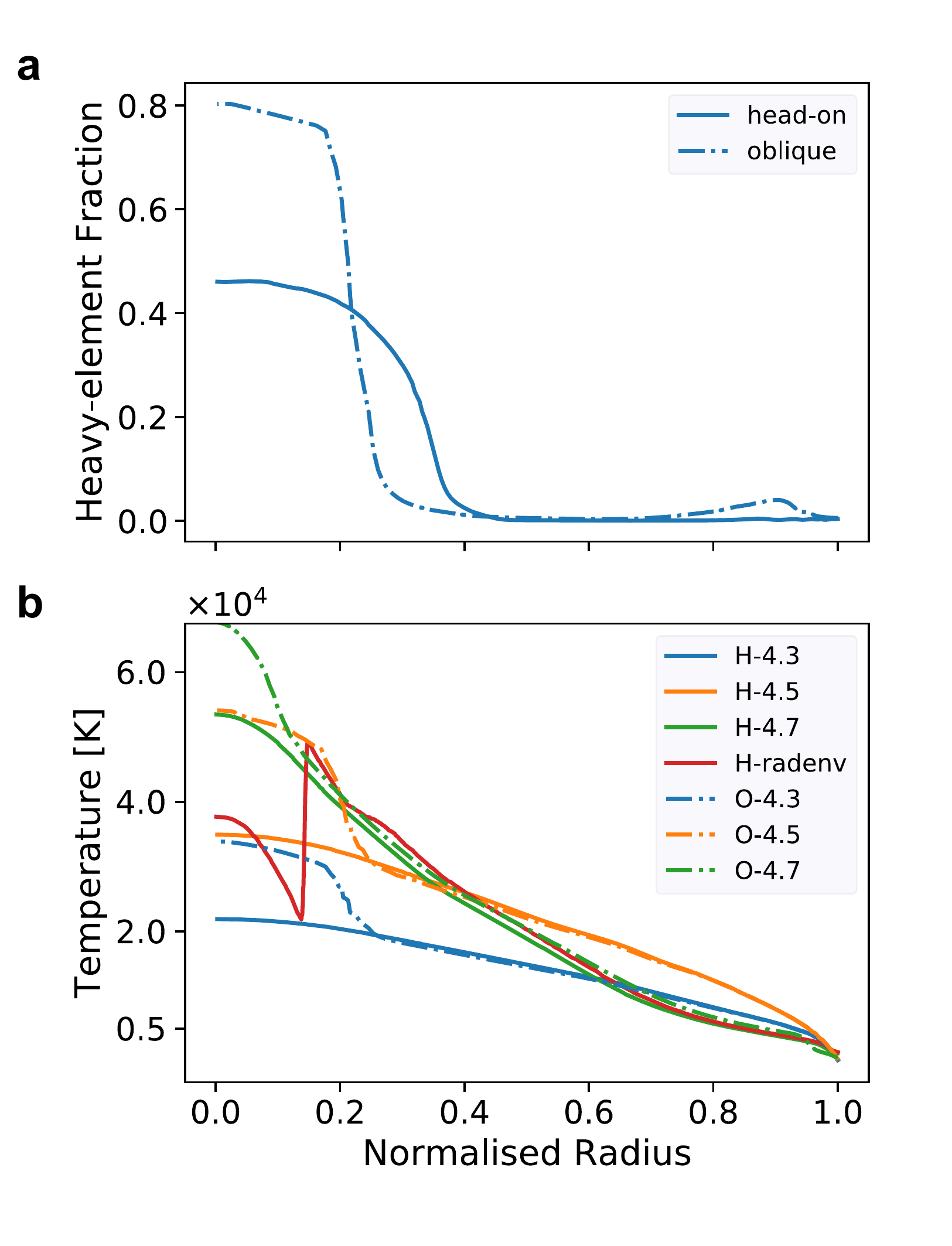}
    \caption{
\textbf{Extended Data Figure 6 | Initial conditions for post-impact evolution. a,} 
  the initial post-impact heavy-element profile and \textbf{b,} temperature profiles of the models 
  that are used for the thermal evolution. The heavy-element distribution is taken from the hydro 
  simulation ten hours after the giant impact. Solid lines correspond to a head-on collision, while 
  dashed-dotted lines show the result of an oblique collision at a 45 degree angle. The colours depict 
  models with different initial thermal states. See text and Extended Data Table 2 for further details.
    }
  \end{center}
\end{figure}   

\begin{figure}[htbp]
  \begin{center}
    \includegraphics[width=\linewidth,clip=true]{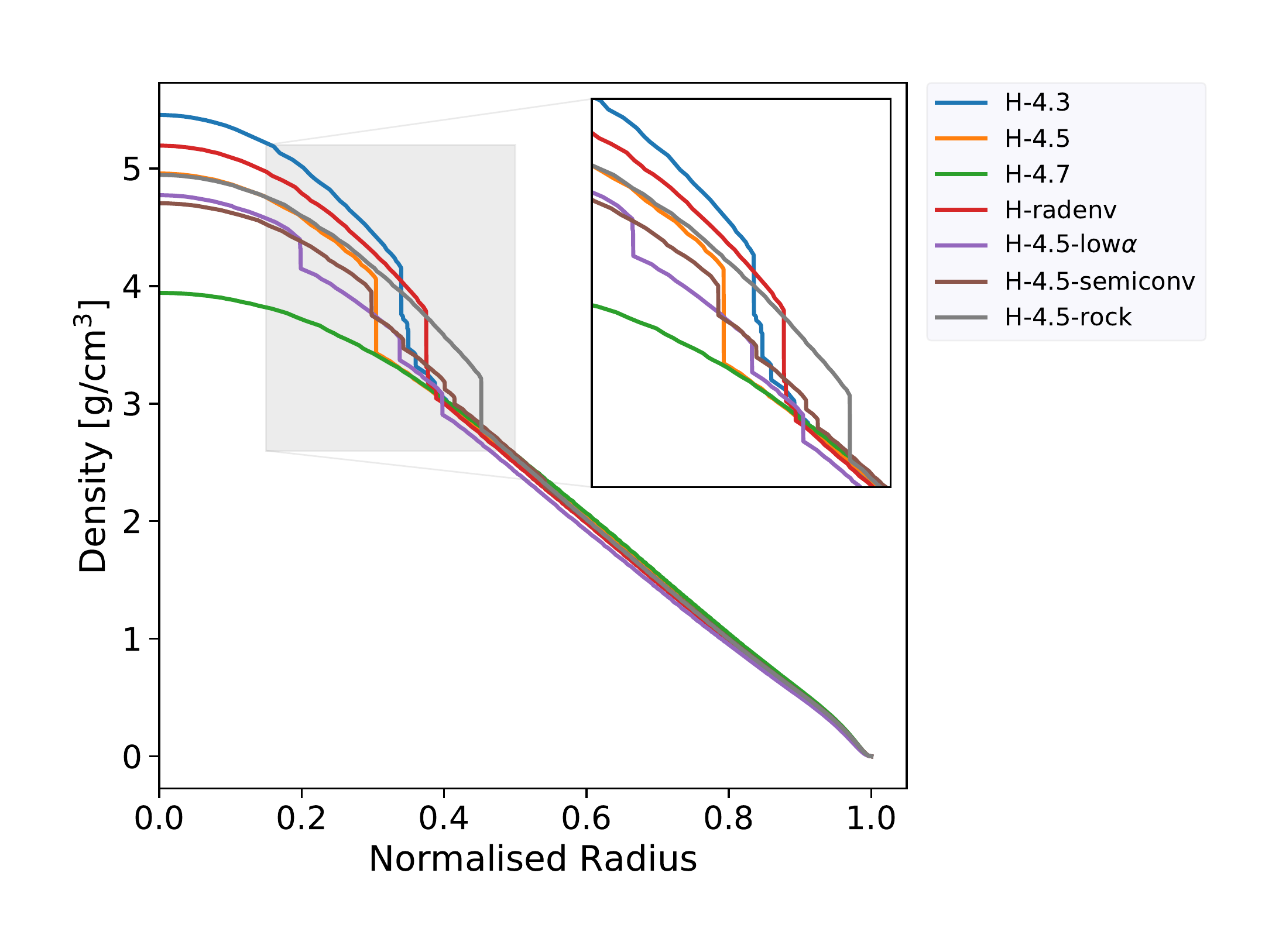}
    \caption{
\textbf{Extended Data Figure 7 | Density vs. normalized radius for the head-on collision 
after 4.56 Gyrs of evolution.} The colors correspond to distinct model assumptions: H-4.3, H-4.5, H-4.7 
correspond to initial thermal profiles with different central temperatures at the time of the impact, while 
H-radenv assumes a proto-Jupiter with a radiative envelope. H-4.5-low$\alpha$ uses a shorter mixing length, 
H-4.5-semiconv allows for semi-convective mixing, and in H-4.5-rock the heavy elements are represented by rock 
instead of water for EOS purposes. See text and Extended Data Table 2 for further details.
    }
  \end{center}
\end{figure}   
   
\begin{figure}[htbp]
  \begin{center}
    \includegraphics[width=\linewidth,clip=true]{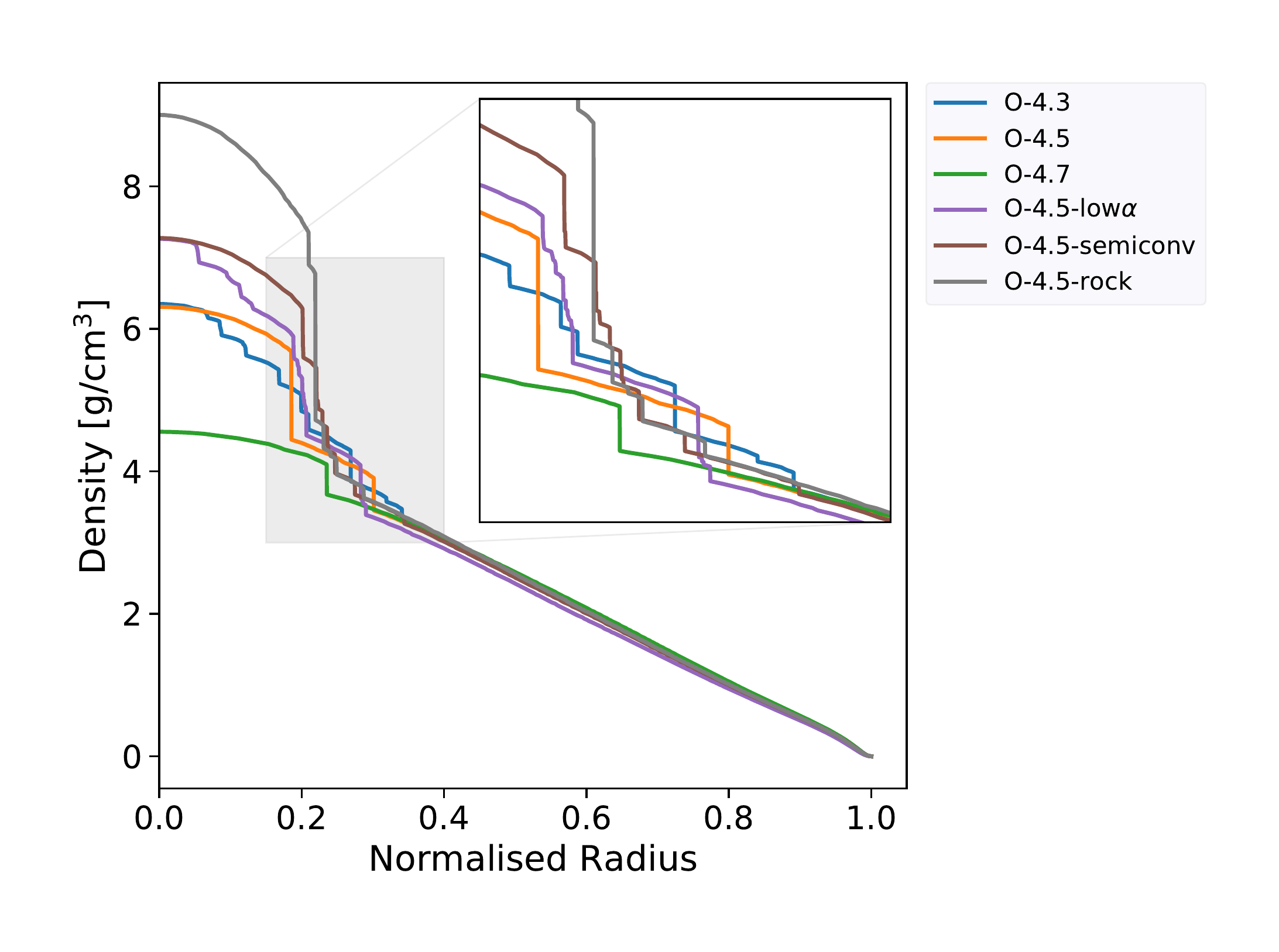}
    \caption{
\textbf{Extended Data Figure 8 | Density vs. normalized radius for the oblique collision 
after 4.56 Gyrs of evolution.} The colors correspond to distinct model assumptions: O-4.3, O-4.5, O-4.7 
correspond to initial thermal profiles with different central temperatures at the time of the impact, 
while O-radenv assumes a proto-Jupiter with a radiative envelope. O-4.5-low$\alpha$ uses a shorter mixing length, 
O-4.5-semiconv allows for semi-convective mixing, and in O-4.5-rock the heavy elements are represented by rock 
instead of water for EOS purposes. See text and Extended Data Table 2 for further details.
    }
  \end{center}
\end{figure}

\newpage 
\begin{figure}[htbp]
  \begin{center}
    \includegraphics[width=\linewidth,clip=true]{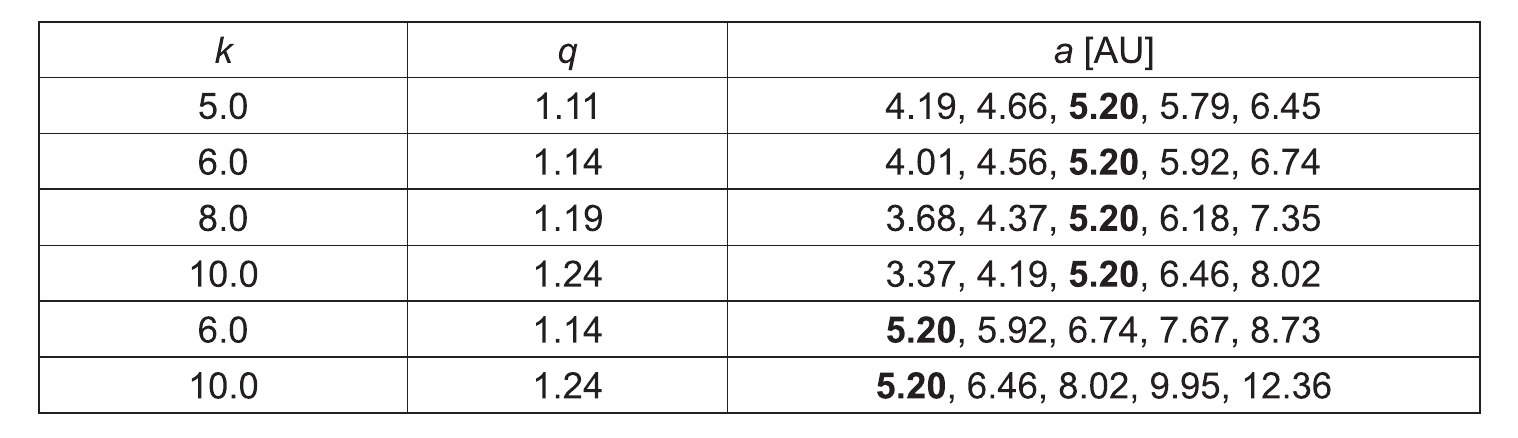}
    \caption{
\textbf{Extended Data Table 1 | List of initial orbital semi-major axis of each embryo of our \textit{N}-body 
    simulation suite.} The location of the embryo that grows into a Jupiter in each case is in bold face. 
    }
  \end{center}
\end{figure}   
    
\begin{figure}[htbp]
  \begin{center}
    \includegraphics[width=\linewidth,clip=true]{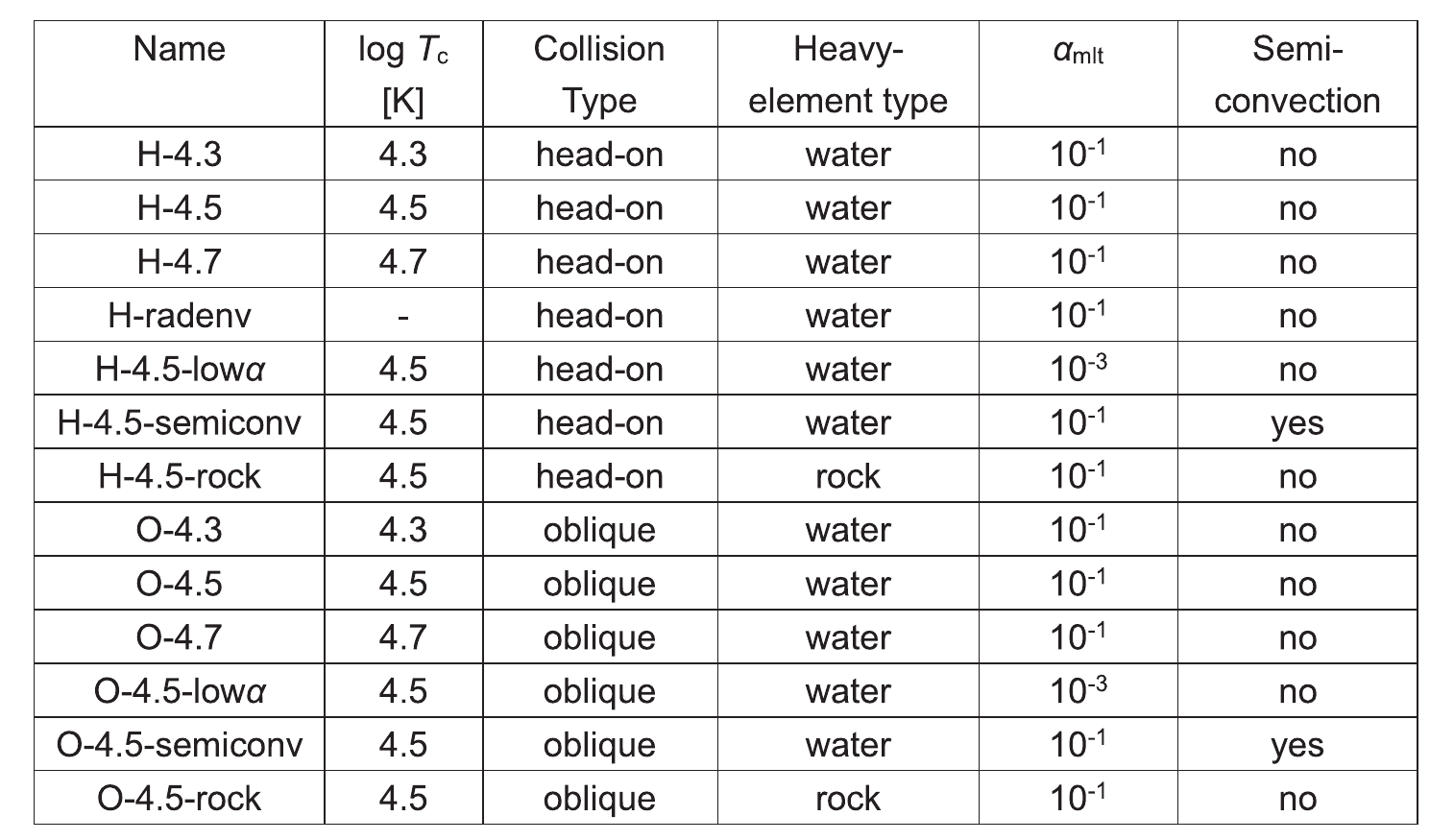}
    \caption{  
\textbf{Extended Data Table 2 | Description of the evolutionary models that are discussed 
    throughout this work.} Note that the models H/O-radenv are unique in which they are the result of formation 
    models of Jupiter\cite{Berardo2017} that account for the accretion shock during the runaway gas accretion.
    }
  \end{center}
\end{figure}

\end{document}